\documentclass{article}
\usepackage{spconf, amsmath, amssymb, siunitx}
\usepackage{graphicx, xcolor, hyperref}

\newcommand{\MAPE}{\mathsf{MAPE}}
\newcommand{\HR}{\mathsf{HR}}
\newcommand{\ECG}{\boldsymbol{\mathsf{ECG}}}
\newcommand{\fakeECG}{\widehat{\ECG}}
\newcommand{\GAN}{\mathsf{GAN}}
\newcommand{\fGAN}{\mathsf{f}\text{-}\mathsf{GAN}}
\newcommand{\freq}{\mathsf{freq}}
\newcommand{\btheta}{\boldsymbol{\theta}}
\newcommand{\bx}{\boldsymbol{x}}
\newcommand{\by}{\boldsymbol{y}}
\newcommand{\bz}{\boldsymbol{z}}
\newcommand{\bX}{\boldsymbol{X}}
\newcommand{\bY}{\boldsymbol{Y}}
\newcommand{\R}{\mathbb{R}}
\newcommand{\E}{\mathbb{E}}
\newcommand{\F}{\mathcal{F}}
\renewcommand{\L}{\mathcal{L}}
\renewcommand{\P}{\mathsf{P}}

\title{$\fGAN$: a frequency-domain-constrained generative adversarial network for PPG to ECG synthesis}
\name{
\begin{tabular}{@{}c@{}}
    Nathan C. L. Kong$^{\dagger * 1}$\thanks{$^1$Work was performed during an internship at Amazon.} \qquad 
    Dae Lee$^*$ \qquad 
    Huyen Do$^*$ \qquad 
    Dae Hoon Park$^*$ \qquad \\
    Cong Xu$^*$ \qquad 
    Hongda Mao$^*$ \qquad 
    Jonathan Chung$^*$
\end{tabular}
}
\address{
    $^\dagger$Stanford University \qquad $^*$Amazon
}

\begin{document}
\ninept

\maketitle


\begin{abstract}
Electrocardiograms (ECGs) and photoplethysmograms (PPGs) are generally used to monitor an individual's cardiovascular health.
In clinical settings, ECGs and fingertip PPGs are the main signals used for assessing cardiovascular health, but the equipment necessary for their collection precludes their use in daily monitoring.
Although PPGs obtained from wrist-worn devices are susceptible to noise due to motion, they have been widely used to continuously monitor cardiovascular health because of their convenience.
Therefore, we would like to combine the ease with which PPGs can be collected with the information that ECGs provide about cardiovascular health by developing models to synthesize ECG signals from paired PPG signals.
We tackled this problem using generative adversarial networks (GANs) and found that models trained using the original GAN formulations can be successfully used to synthesize ECG signals from which heart rate can be extracted using standard signal processing pipelines.
Incorporating a frequency-domain constraint to model training improved the stability of model performance and also the performance on heart rate estimation.
\end{abstract}


\section{Introduction}
\label{sec:intro}
There is an increasing interest in cardiovascular wellness \cite{USPS_JAMA, Javed2022} and there exists various non-invasive approaches for monitoring cardiovascular health.
Clinically, electrocardiograms (ECGs) and fingertip photoplethysmograms (PPGs) are used along with established signal processing pipelines for extracting clinically-validated traits.
Continuous monitoring precludes the use of them due to the equipment necessary for their collection.
The increasing prevalence of wrist-worn devices may alleviate this problem, since PPGs are easily collected by these devices \cite{Charlton2022Wear}.

There are trade-offs to consider when choosing between ECGs or PPGs.
Each signal modality differs in their availability, signal quality and the mode of measurements.
For example, ECGs are less noisy with higher sampling rate because they are usually collected in stationary settings.
In addition, ECGs monitor cardiac electrical activity whereas PPGs monitor blood flow in the periphery using optical sensors.
Therefore, it is of interest to combine the richness of ECGs with the availability of PPGs.

We work towards this goal by synthesizing ECG signals directly from (paired) PPG signals with a few desiderata.
First, we would like a model that can synthesize ECG signals from PPG signals that are obtained from \emph{wearable} devices during daily activities, as opposed to from PPG signals obtained using fingertip pulse oximeters.
This is because wearable devices more readily allow for the continuous monitoring of biometric signals during typical daily activities.
Second, we would like a PPG-to-ECG synthesis model that can generalize across subjects so that it can be more readily used off-the-shelf.

Here, we used the framework of generative adversarial networks (GANs, \cite{Goodfellow2014}) to learn the mapping from PPG signals to ECG signals.
We did not pose this problem as a supervised translation problem (e.g., Euclidean distance minimization) because GANs allow us to not only learn the mapping function (from PPG to ECG), but also remove the need to hand-engineer the objective function for synthesizing realistic data samples \cite{Isola2017}.
Further motivating our use of GANs is their successful applications in image-to-image translation problems \cite{Isola2017, Zhu2017}.

There are a few related works that used GANs for synthesizing ECG signals from PPG signals.
Using the MIMIC dataset \cite{Johnson2016, Goldberger2000, Moody1996}, it has been shown that ECG signals could be generated with high-fidelity from PPG signals obtained via fingertip pulse oximetry \cite{Vo2021}.
Other work showing successful PPG-to-ECG synthesis used a GAN that incorporated both time-domain and frequency-domain inputs along with constraints on backwards translation (i.e., ECG-to-PPG) \cite{Sarkar2021}.
Their model was very sophisticated, but it was unclear which components contributed the greatest to model performance and how stable it was across random initializations.
Finally, models that do not use GANs have also been shown to be capable of PPG-to-ECG reconstruction, but are subject-specific and are built using PPG signals obtained from fingertip pulse oximetry \cite{Tang2022}.

Our work focusses on building models to synthesize ECGs from PPGs using GANs starting with the most basic formulation of them and on investigating the benefits that certain model components confer.
We found that training a model using the original GAN formulation resulted in synthetic ECG signals that can be used for heart rate estimation, serving as a good baseline.
Furthermore, adding a frequency-domain constraint during model training improves the properties of model training and also improves model performance on heart rate estimation.

\section{Methodology}
\label{sec:methodology}

\subsection{Dataset and Signal Preprocessing}
\label{ssec:dataset}
We used the dataset collected by Reiss et al. \cite{Reiss2019}, known as \emph{PPG-Dalia}.
It consists of a set of synchronized PPG and ECG signals obtained from $15$ participants while they performed a wide variety of natural activities, including sitting (at rest), cycling, working, walking and playing table soccer, etc., over a span of approximately two hours.
PPG signals in this dataset were collected from a wrist-worn device at $64$ Hz, while ECG signals were (simultaneously) collected at $700$ Hz using a chest-worn device.

We next performed basic signal preprocessing of each participant's data.
The PPG and the ECG signals were first resampled to $128$ Hz.
They were then segmented into overlapping four-second windows (i.e., $512$ time points for each data sample), where adjacent data samples had a two-second overlap.
After signal segmentation, a bandpass filter was applied on each signal using a Python package known as \texttt{biosppy} \cite{biosppy}.
Each PPG segment was bandpass filtered using a fourth order Chebyshev Type II filter with passband frequencies of $0.4$~Hz and $8$~Hz.
Similarly, each ECG segment was bandpass filtered using a finite impulse response (FIR) filter with passband frequencies of $3$~Hz and $45$~Hz.
Finally, both signal segments were min-max scaled to $[-1,1]$.
These preprocessing steps are similar to those used in prior work \cite{Sarkar2021}.

Prior to model training and evaluation, the data were split into train, validation and test sets.
Nine random participants were assigned to the train set ($\num{40675}$ segments), three random participants were assigned to the validation set ($\num{11276}$ segments) and the final three participants were assigned to the test set ($\num{12796}$ segments).
All three sets were disjoint and consisted of data from \emph{different} subjects, so that our models are \emph{subject agnostic}.

\subsection{Generative Adversarial Networks and Objective Functions}
\label{ssec:gans}
At a high-level, GANs are based on a two-player game, where both players can be deep neural networks.
One player is the ``generator'' ($G$) and its goal is to produce synthetic data (e.g., ECG signals) that are as indistinguishable as possible from real data.
The other player is the ``discriminator'' ($D$) and its goal is to determine whether or not its input is synthetic (i.e., fake).
Thus, the generator and the discriminator have opposing objectives and are trained \emph{adversarially} until they reach equilibrium.

We experimented with two different objective functions when training a generator to synthesize ECG signals from PPG signals.
The first objective function was the original adversarial loss formulation \cite{Goodfellow2014}, defined as follows:
\begin{equation}
\begin{aligned}
    \L_{\GAN}(\bX, \bY; \btheta_G, \btheta_D) = \E_{\bx \sim \P_{\bX}(\bx)}[\log (1 - D(G(\bx))] \, + \\
    \E_{\by \sim \P_{\bY}(\by)}[\log D(\by)],
\end{aligned}
\label{eq:original-gan}
\end{equation}
where $\btheta_G$ and $\btheta_D$ are the parameters of the generator and the discriminator respectively, $\bX$ is the set of PPG signals, $\bY$ is the set of ECG signals, $\bx \in \R^{512}$ is a PPG signal and $\by \in \R^{512}$ is a real ECG signal.
Since the objective of $D$ is to distinguish between synthetic and real ECG signals, $D$ aims to maximize $\L_{\GAN}$.
On the contrary, $G$ aims to synthesize ECG signals so real as to fool $D$.
Thus, $G$ aims to minimize $\L_{\GAN}$.

The second objective function builds upon Equation~\ref{eq:original-gan} by incorporating a constraint in the frequency domain, since constraints in the time-domain may not be able to handle the undesirable effects due to the different latencies between PPG peaks and ECG peaks across participants.
The constraint is also motivated by the desire to encourage greater morphological similarity between the synthetic and the real ECG signals (e.g., number of R-peaks, P-waves, T-waves).
Concretely, the frequency-domain constraint is defined as follows:
\begin{equation}
    \L_{\freq}(\bX, \bY; \btheta_G) = \E_{\bx, \by \sim \P_{\bX, \bY}(\bx, \by)} \left[ \left\lVert \, \lvert \F(G(\bx)) \rvert - \lvert \F(\by) \rvert \, \right\rVert_1 \right],
\end{equation}
where $\lvert \F(\bz) \rvert$ denotes the amplitude of each frequency component of $\bz$, excluding the $0$ Hz component (i.e., DC component), obtained via a fast Fourier transform.
The final objective incorporating the frequency-domain constraint is:
\begin{equation}
\begin{aligned}
    \L_{\fGAN}(\bX, \bY; \btheta_G, \btheta_D) = \L_{\GAN}(\bX, \bY; \btheta_G, \btheta_D) \, + \\
    \lambda_{\freq} \L_{\freq}(\bX, \bY; \btheta_G),
\end{aligned}
\label{eq:freq-orig-gan}
\end{equation}
where $\lambda_{\freq}$ is the coefficient that weights the relative importance of the frequency-domain constraint.
This combined objective therefore encourages the generator to both fool the discriminator while remaining close to the real data in the frequency domain in an $\ell_1$ sense.

\subsection{Model Architecture}
\label{ssec:model-architecture}
\begin{figure}[htb]
    \centering
    \includegraphics[width=0.96\linewidth]{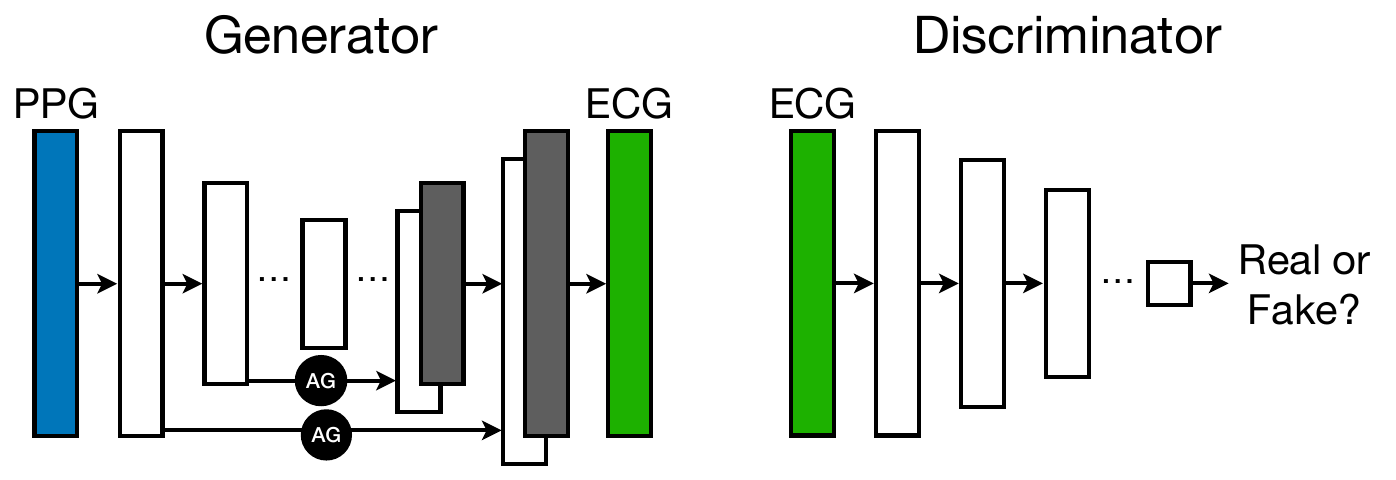}
    \caption{\textbf{Schematics of the generator and the discriminator architectures.}
    ``AG'' denotes the attention gate when combining features from the encoder with the features from the decoder.
    Left: architecture of the generator, which takes as input the PPG signal.
    Right: architecture of the discriminator, which takes as input the real or the synthetic ECG signal.
    }
    \label{fig:architecture}
\end{figure}

The generator architecture was a U-Net with skip connections \cite{Ronneberger2015, Isola2017, Sarkar2021}.
In addition to the U-Net structure, we applied attention gates at the output of each skip connection so that the model learns to emphasize input features at various resolutions that are useful for a task, as in prior work \cite{Oktay2018, Sarkar2021}.
The ``encoding'' portion of the generator consisted of six 1D convolutional layers with an increasing number of output filters: $(64, 128, 256, 512, 512, 512)$.
The first three convolutional operations had a stride of two and the latter three convolutional operations had a stride of one.
The ``decoding'' portion of the generator was essentially the ``mirrored'' version of the encoder, where spatial size is gradually increased by sequentially applying upsampling and convolutional operations.
All the convolutional filters had a kernel size of $16$.
A high-level schematic of the generator architecture is shown on the left of Figure~\ref{fig:architecture}.

The discriminator architecture was a six-layer 1D convolutional neural network.
The number of output filters for each layer increased as a function of depth: $(32, 64, 128, 256, 512)$.
Another 1D convolutional layer was applied on the output of these five convolutional layers to reduce the number of output channels from $512$ to one.
Finally, this output was reduced to a scalar value via an average-pooling operation (the scalar value is used to determine the probability that the input is real or fake).
All the convolutional filters had a kernel size of $16$ with a stride of one.
A high-level schematic of the discriminator architecture is shown on the right of Figure~\ref{fig:architecture}.

\subsection{Model Training}
\label{ssec:model-training}
The objective functions described in Equations~\ref{eq:original-gan} and \ref{eq:freq-orig-gan} were optimized using the Adam optimizer \cite{Kingma2015} with a learning rate of $10^{-5}$ for the discriminator, a learning rate of $10^{-4}$ for the generator and a batch size of $128$.
The beta coefficients for the optimizer were set to $\beta_1 = 0.9$ and $\beta_2 = 0.999$.
The discriminator was also updated five times slower than the generator (i.e., the discriminator is updated every five training iterations and the generator is updated every iteration).
To optimize Equation~\ref{eq:original-gan}, the models were trained for $15$ epochs with the learning rates constant for four epochs and then linearly decayed to zero.
To optimize Equation~\ref{eq:freq-orig-gan}, $\lambda_{\freq}$ was set to $0.1$ and the models were trained for $11$ epochs with the learning rates constant for five epochs and then linearly decayed to zero.
Finally, the optimization of each objective function was performed $31$ times, each time with a different random seed.
All models and optimization were implemented using PyTorch.

\subsection{Model Evaluation}
\label{ssec:model-evaluation}
We evaluated each synthetic ECG signal based on how well heart rate could be estimated from the signal with respect to the heart rate estimated from the real ECG signal, as it is more interpretable than other metrics such as root mean-squared error.
A $10$-second ECG segment allows for heart rate to be extracted more reliably.
Therefore, we first segmented the signals from the validation and the test sets into $10$-second paired PPG and ECG segments (with eight-second overlaps between consecutive samples).
$10$-second synthetic ECG signals were then generated for each $10$-second PPG signal.
A popular peak detection algorithm \cite{Hamilton2002, biosppy} was applied to extract heart rate from both the synthetic and the real ECG signals, same as that used in prior work \cite{Reiss2019, Sarkar2021}.
We then computed the absolute difference between the two estimated heart rates scaled by the heart rate of the real ECG signal.
Concretely, the mean absolute percentage error (denoted as $\MAPE$) across the dataset is defined as follows:
\begin{equation}
    \MAPE(\ECG, \fakeECG) = \frac{100}{N} \sum_{i=1}^N \frac{\lvert \HR(\ECG_i) - \HR(\fakeECG_i) \rvert}{\HR(\ECG_i)},
\label{eq:hr-est-error}
\end{equation}
where $N$ is the number of (PPG or ECG) signal segments in either the validation or the test set, $\ECG_i$ is the $i$th real ECG signal, $\fakeECG_i$ is the $i$th synthetic ECG signal and $\HR(\cdot)$ is the peak detection algorithm used to estimate heart rate from ECG signals \cite{Hamilton2002}.

\section{Results}

\subsection{Qualitative Results}
\begin{figure}[htb!]
    \centering
    \includegraphics[width=0.98\linewidth]{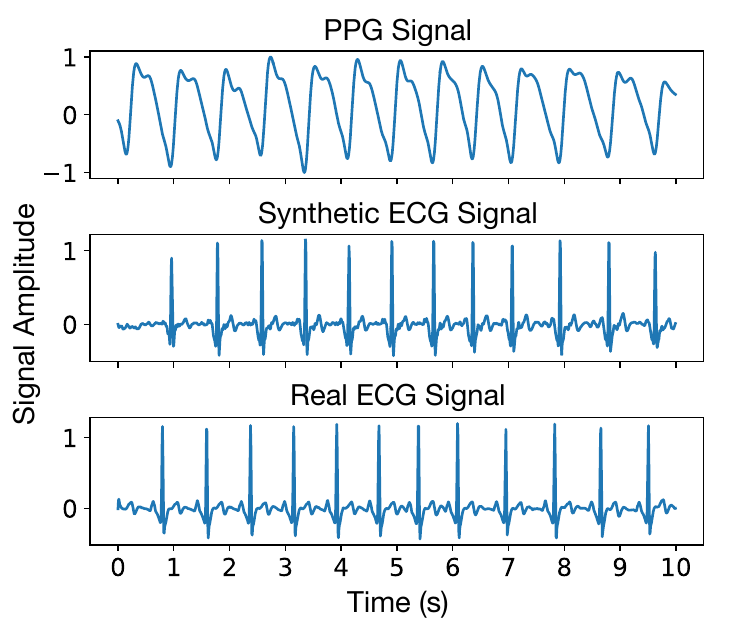}
    \caption{\textbf{Example of a synthesized ECG signal from a PPG signal using a GAN trained with the frequency-domain constraint.}
    The PPG signal (top) was used as the input to the generator, which output a synthetic ECG signal (middle).
    The synthetic ECG signal looks qualitatively similar to the real (paired) ECG signal.
    Signal amplitudes shown here were obtained \emph{after} signal preprocessing steps were performed.
    }
    \label{fig:vanilla_freq_example}
\end{figure}

Figure~\ref{fig:vanilla_freq_example} shows qualitatively that a GAN trained with the frequency-domain constraint (Equation~\ref{eq:freq-orig-gan}) can synthesize ECG signals that look very similar to the real ECG signal.
Furthermore, we can observe some aspects of ECG-signal morphology, including the P-wave, the QRS complex and the T-wave, in each heartbeat.

\subsection{Improved Stability of Model Performance Across Random Seeds}
\begin{figure}[htbp]
    \centering
    \includegraphics[width=0.98\linewidth]{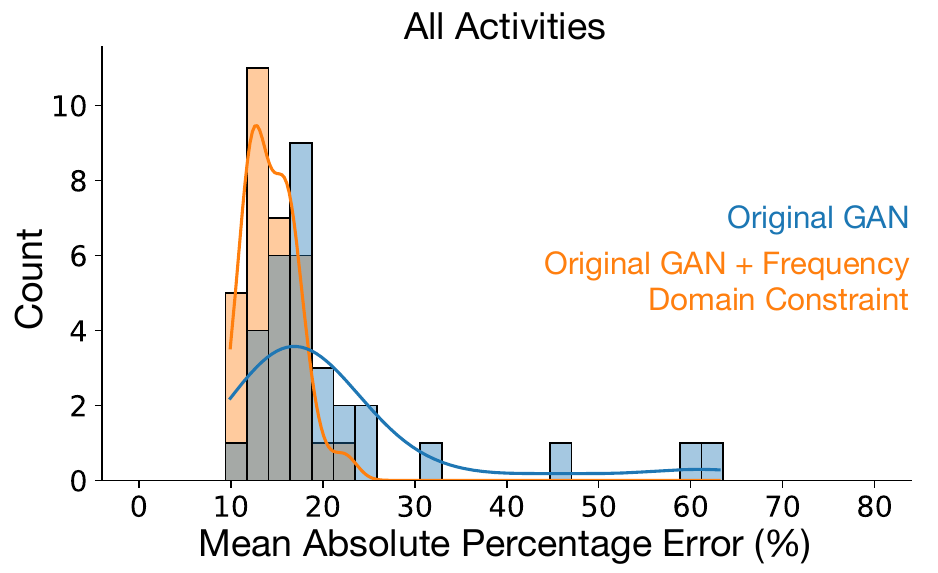}
    \caption{\textbf{Distribution across random seeds of model performance on heart rate estimation during all activities for the two objective functions.}
    Across the entire validation set, the model incorporating the frequency-domain constraint performs better (on average across seeds) than the model without the constraint ($t(60) = 3.04, p = 0.005$) and its performance is less variable across random seeds.
    }
    \label{fig:hr_est_error_all}
\end{figure}

GANs are known to have non-convergence issues \cite{Salimans2016, Goodfellow2016}.
We therefore ascertained the stability of each model's performance on heart rate estimation across $31$ random seeds (i.e., random initializations and dataset shuffles) using the validation set.
The distribution of the mean absolute percentage error for both objective functions across the random seeds is shown in Figure~\ref{fig:hr_est_error_all}.
We found that across the random seeds, model performance was more variable if a frequency-domain constraint was not incorporated, as can be seen by the heavily right-skewed distribution in Figure~\ref{fig:hr_est_error_all}.
Specifically, the standard deviation in model performance across random seeds for the model trained without the frequency-domain constraint was $12\%$ and was $3\%$ for the model trained with the constraint.
Furthermore, the average of the mean absolute percentage error across random seeds was smaller for the model trained with the constraint ($p = 0.005$).

\begin{figure}[!ht]
    \centering
    \includegraphics[width=0.98\linewidth]{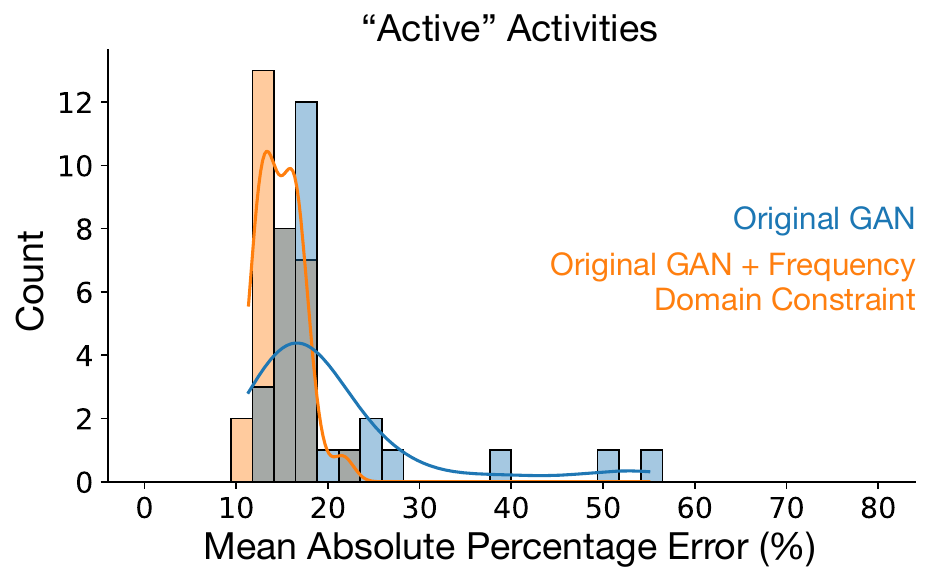}
    \caption{\textbf{Distribution across random seeds of model performance on heart rate estimation during ``active'' activities for the two objective functions.}
    When evaluated on signal segments during staircase traversal, table soccer, cycling, driving, walking and working, the model incorporating the frequency-domain constraint has better performance (on average across seeds) than the model without the constraint ($t(60) = 2.98, p = 0.005$).
    Incorporating the constraint also reduces performance variance.
    }
    \label{fig:hr_est_error_active}
\end{figure}

We subset the validation set into segments that were associated with ``more active'' activities to investigate how the models performed under activities that have higher heart rates.
These activities consisted of going up and down stairs, playing table soccer, cycling, driving, walking and working.
The performance distributions evaluated over the ``active'' activities shifted slightly to the right (compare the positions of the leftmost bars in Figures~\ref{fig:hr_est_error_all} and \ref{fig:hr_est_error_active}).
Overall, the observations of improved average model performance and reduced performance variance across random seeds is consistent across an evaluation using a subset of activities containing higher heart rates.

\subsection{Reduction in the Number of Heart Rate Estimation Failure Cases}
\begin{figure}[ht]
    \centering
    \includegraphics[width=0.98\linewidth]{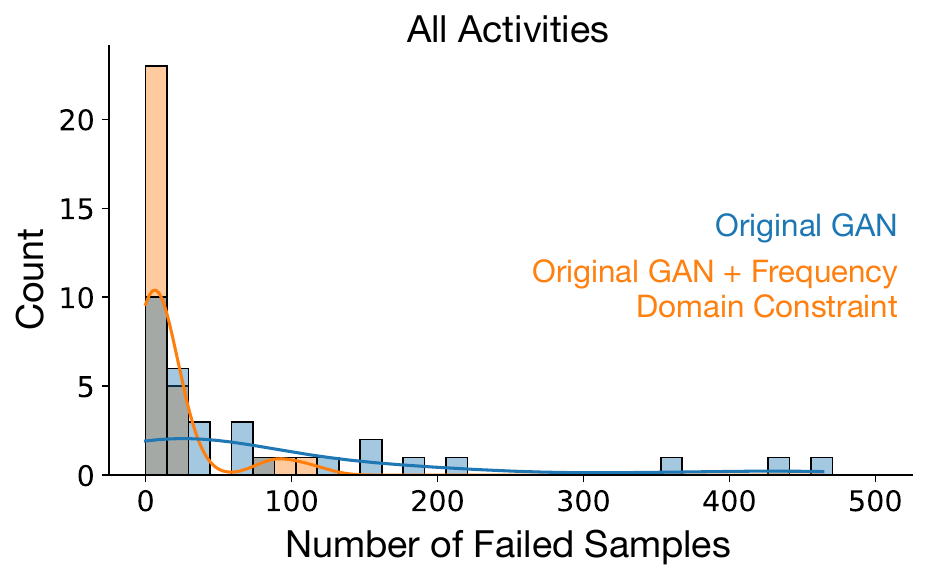}
    \caption{\textbf{Distribution across random seeds of total number of samples where heart rate estimation failed during all activities for the two objective functions.}
    Across the entire validation set (a total of $\num{11276}$ samples), the model incorporating the frequency-domain constraint results in less samples in which heart rate estimation fails (on average across seeds) than the model without the constraint ($t(60) = 2.97, p = 0.006$), while using existing signal processing pipelines.
    }
    \label{fig:hr_total_failed_all}
\end{figure}

One important use-case of the synthetic ECG signals is the ability for them to be able to be incorporated into well-established signal processing pipelines that can detect different morphological properties of ECGs.
To assess a model's ability for this use-case, we computed, for each model (and for each random seed), the total number of validation set samples in which a standard peak-detection algorithm \cite{Hamilton2002, biosppy} failed to detect heart rate.
Across the random seeds, we found that the model trained without the frequency-domain constraint had many more failure cases than the model trained with the constraint, shown by the long right tail of the ``Original GAN'' distribution in Figure~\ref{fig:hr_total_failed_all} ($p = 0.006$).
Thus, training models using an additional frequency-domain constraint more readily allows for existing ECG signal processing pipelines to be used.

\subsection{Reduction in Average Heart Rate Estimation Error}
We next evaluated the best performing model (chosen according to the validation set), for each objective function, on the test set.
We also compared the models' performance with respect to the mean absolute percentage error obtained if the peak-detection algorithm of Elgendi et al. \cite{Elgendi2013, Makowski2021} were used directly on the PPG signals, without accelerometer data.
We found that the model trained with the frequency-domain constraint outperforms the model trained without the constraint by $2\%$ when all the activities are considered (leftmost column of Table~\ref{tab:test-set-results}).
It also outperforms a strong PPG peak-detection algorithm for heart rate estimation by $3\%$.

\begin{table}[!ht]
\centering
\begin{tabular}{|c||c|c|c|}
\hline
 & All & Not Active & Active \\ \hline\hline
PPG \cite{Elgendi2013, Makowski2021} & $15\%$ & $14\%$ & $17\%$ \\ \hline
\begin{tabular}[c]{@{}c@{}}Original GAN\\ (Equation~\ref{eq:original-gan})\end{tabular} & $14\%$ & $14\%$ & $14\%$ \\ \hline
\begin{tabular}[c]{@{}c@{}}Original GAN + \\ Frequency-Domain \\ Constraint (Equation~\ref{eq:freq-orig-gan})\end{tabular} & \underline{$\mathbf{12\%}$} & \underline{$\mathbf{12\%}$} & \underline{$\mathbf{12\%}$} \\ \hline
\end{tabular}
\caption{\textbf{Mean absolute percentage error on the test set (a total of $\num{12796}$ samples) for different activity subsets using the best model determined by the validation set.}
For the PPG signals, heart rate was estimated using the peak-detection algorithm of Elgendi et al. \cite{Elgendi2013, Makowski2021} and the error was computed with respect to the heart rate estimated from the paired ECG signal.
}
\label{tab:test-set-results}
\end{table}

\section{Conclusions}
The increasing adoption of health-oriented wearable devices will increase the availability of PPG signals.
A subject-agnostic model to generate ECG signals from PPG signals could provide vast amounts of information of wearable-users' cardiovascular health.
This information could potentially be used as a continuous-monitoring platform for existing cardiovascular conditions.

In this work, we have made some additional progress into this problem by providing a basis upon which next-generation, GAN-based models could be developed.
Firstly, by converting PPG signals to ECG signals, we obtained more accurate heart rate estimations when compared to state-of-the-art PPG heart rate estimation algorithms.
This was especially prominent when participants were performing activities causing noisier PPG signals.

We also showed that incorporating a frequency-domain constraint during subject-agnostic model training conferred several advantages.
It led to improved performance stability (across random seeds) during GAN training, measured by average heart rate estimation error.
Moreover, the constraint led to a reduced number of heart rate estimation failure cases, therefore improving the reliability of the detected heart rates so that they can be more readily used with existing ECG signal processing pipelines.
Finally, models trained with the constraint had reduced average heart rate estimation error compared to the model trained without the constraint.

\vfill\pagebreak
\bibliographystyle{IEEEbib}
\bibliography{main}

\end{document}